\documentclass[twocolumn,10pt, aip]{revtex4}
\usepackage{graphicx}
\usepackage{wrapfig}
\usepackage{dcolumn}
\usepackage{bm}

\begin{document}
\newcommand{\bR}{{\bf R}}

\title{Energetics and Dipole Moment of Transition Metal Monoxides by Quantum Monte Carlo}
\author{Lucas K. Wagner, Lubos Mitas}

\affiliation{ Center for High Performance Simulation and
Department of Physics,
North Carolina State University, Raleigh, NC 27695.}

\date{\today}

\begin{abstract}

The transition metal-oxygen bond appears very prominently throughout chemistry and solid-state physics.  Many materials, 
from biomolecules to ferroelectrics to the components of supernova remnants contain this bond in some form.  Many of these materials' properties depend strongly on fine details of the TM-O bond, which makes accurate calculations of their properties
 very challenging.  Here we report on highly accurate first principles calculations of the properties of TM monoxide molecules within fixed-node Diffusion Monte Carlo and Reptation Monte Carlo.
\end{abstract}


\maketitle

\section{Introduction}

Transition metal chemistry is an exciting area of research that has implications in fields from biological physics to astrophysics.  Transition metals can form many types of bonds and transition metal solids exhibit useful properties like ferroelectric and ferromagnetic ordering.  This interesting physics comes from the $d$-shell states which are very close in
energy to the outer $s-$states, along with strongly correlated electrons that make accurate first principles calculations particularly challenging.  
Benchmark calculations are particularly useful to determine what level of accuracy one can obtain from a given method, although
the precise bonding pattern can vary from system to system.

Many authors (most recently, Refs \cite{furche:044103,baushlicher:189}) have studied the transition metal monoxides using Density Functional Theory and post-Hartree-Fock methods.  The performance of these methods is less
reliable whenever transition metals are included in a system.
In particular,  the calculation of dipole moments is challenging because
it is rather sensitive to the details of calculations and sizes of
employed basis sets. The results can sometimes be far from experiment.  The TM-O bond is the driving force behind many perovskite and
earth materials, 
 which have been noted as having significant errors in the unit cell volume within DFT\cite{Tinte,Korotin}, while being too large for
post Hartree-Fock methods to be applied.
 We have had some success\cite{CPL_lucas} applying ground-state quantum Monte Carlo (QMC) to the binding
energies of the TiO and MnO molecules, which hinted that QMC may be able to treat these systems more accurately.  QMC also has the property of scaling well with system size, although with a large 
prefactor, and has seen limited application to TM solids.  As more computing power becomes available, QMC calculations of solids will become routine, and this study offers some insight as to the accuracy that will be achieved.

In this paper, we expand our treatment to the first five TM-O molecules (Sc,Ti,V,Cr,and Mn), studying not only the binding energy, but also the bond length and the dipole moment.  To obtain the dipole moment, we apply the relatively new Reptation Monte Carlo\cite{Baroni_RMC} method for the first time to heavy elements.  We find that for binding energy and bond lengths, QMC offers unmatched
accuracy, while the dipole moment is in less agreement with experiment.
 We investigate the effect of the fixed-node condition 
on the dipole moment and find that while there 
is a significant nodal error, 
it is not enough to reconcile the calculation with the experiments. 

\section{Method}

\subsection{Quantum Monte Carlo}

We use the Variational, Diffusion, and Reptation flavors of Quantum Monte Carlo (VMC,DMC, and RMC) in our calculations as implemented in the QWalk program\cite{qwalk}.  
In VMC, we start with a Slater determinant of one-particle orbitals, $\Psi_{HF}$, or a linear combination of Slater determinants.  We then multiply $\Psi_{HF}$ by the explicitly correlated inhomogeneous Jastrow correlation factor $e^U$.  We write 
\begin{equation}
U=  \sum_{ijI} u(r_{iI},r_{jI},r_{ij})
\end{equation}
where the lower case indices stand for electronic coordinates, and the upper case indices are ionic
coordinates.  The correlation factor is expanded in the Schmidt-Moskowitz form\cite{schmidt:4172}:
\begin{eqnarray*}
u(r_{iI},r_{jI},r_{ij})=\sum_k c_k^{ei}a_k(r_{iI}) + \sum_m c_m^{ee} b_k(r_{ij}) \\
+ \sum_{klm} c_{klm}^{eei} (a_k(r_{iI})a_l(r_{jI})+a_k(r_{jI})a_l(r_{iI}))b_k(r_{ij}),
\end{eqnarray*}
where the $a_k$ and $b_k$ functions are written as 
\begin{equation}
a_k(r) = \frac{1-z(r/r_{cut})}{1+\beta z(r/r_{cut})}.
\end{equation}  The polynomial
$z(x)=x^2(6-8x+3x^2)$ is chosen so the functions go smoothly to zero at $r_{cut}=$7.5 bohr.  
We generate random samples in the $3N_e$-dimensional space (denoted by $\bR$) according to the many-particle probability distribution $\Psi(\bR)^2$.  The energy is then obtained by averaging the local energy $E_L(\bR)=\frac{H\Psi(\bR)}{\Psi(\bR)}$.  Following the variational theorem,
we then optimize a combination of energy and variance of the local
energy, using the algorithm proposed by Umrigar and Filippi\cite{umrigar_optimization2}.
$\beta$ and all the coefficients are variationally 
optimized.
We use the VMC wave function as a trial function for Reptation Monte Carlo or
Diffusion Monte Carlo.  

DMC and RMC are based on the so-called imaginary time Schr\"odinger equation
\begin{equation}
-\frac{d\Psi(\bR,\tau)}{d\tau}=(H-E_0)\Psi(\bR,\tau),
\end{equation}
which has a steady-state solution when $\Psi$ is an eigenvalue with value $E_0$ and all 
non-steady-state solutions converge exponentially to the eigenstate $\Phi_0$ as $\tau$ 
goes to infinity.  Transforming to an integral equation, we have
\begin{equation}
\Phi_0(\bR_1)=\lim_{\tau \to \infty} \int G(\bR_1,\bR_0,\tau) \Psi_T(\bR_0) d\bR_0
\label{eqn:imagtime}
\end{equation}
where $G$ is the Green's function of the imaginary time Schr\"odinger equation and $\Psi_T(\bR_0)$
is the trial wave function that we obtain from VMC.
Solving for the exact $G$ for large $\tau$ is as difficult as solving for $\Phi_0$, so we choose
some constant small value of $\tau$ for which we know $G$ accurately, and compound the operations
(suppressing the $\tau$ dependence of $G$):
\begin{equation}
\Phi_0(\bR)=\lim_{n \to \infty} G(\bR,\bR_n)\ldots G(\bR_1,\bR_0)\Psi_T(\bR_0).
\end{equation}
Each application of $G$ is interpreted as a stochastic process, in the same way that the 
diffusion equation can be mapped onto Brownian particles and vice versa (in fact, for a 
free particle, the Hamiltonian is $-\frac{1}{2}\nabla^2$ and the simulation is a diffusion 
process).

 DMC performs a simulation of random particles for large $n$.  Skipping over some details
that can be found in Ref \cite{Foulkes_review}, we eventually find that it obtains 
 $P_{R_\infty}(\bR)=\Phi_0(\bR)\Psi_T(\bR)$, which 
can be used to evaluate the ground-state energy as follows:
\begin{equation}
\langle E_0 \rangle = \int d\bR \Psi_T(\bR)\Phi_0(\bR) \frac{H\Psi_T(\bR)}{\Psi_T(\bR)},
\end{equation}
since $\Phi_0$ is an eigenstate of $H$ and $H$ can operate forwards or backwards.
 Any operators that do not commute with the Hamilonian will have expectation values in error. 
We account for this error by using Reptation Monte Carlo, where the random walk is performed in the space of paths: 
$s=[\bR_0, \bR_1, \ldots,\bR_{n-1}, \bR_n]$.  We sample the path probability distribution
\begin{equation}
\Pi(s)=\Psi_T(\bR_0) G(\bR_0,\bR_1)\ldots G(\bR_{n-1},\bR_n) \Psi_T(\bR_n)
\end{equation}
This can be interpreted in several different ways.  If we examine the distribution at
$\bR_0$, we can view the samples of Green's functions as acting on $\Psi_T(\bR_n)$, 
and therefore $P_{R_0}(\bR_0)=\Psi_T(\bR_0)\Phi_0(\bR_0)$.  This is the same distribution as
we obtain DMC as the path length goes to infinity.  Alternatively, 
since $G$ is symmetric on exchange of the two $\bR$ coordinates, the probability distribution 
of $\bR_n$ is the same.  Finally, we can split the path in two, one projecting on 
$\Psi_T(\bR_0)$, and the other projecting on $\Psi_T(\bR_n)$.  We then have 
\begin{eqnarray*}
P_{R_{n/2}}(\bR_{n/2})=( G(\bR_{n/2},\bR_{n/2-1})\ldots G(\bR_1,\bR_2) \Psi_T(\bR_0) )\\
\times ( G(\bR_{n/2},\bR_{n/2+1})\ldots G(\bR_{n-1},\bR_n) \Psi_T(\bR_n) ) \\
= \Phi_0^2(\bR_{n/2})
\end{eqnarray*}
for $n \rightarrow \infty$, which allows us to obtain correct expectation values of
operators that do not commute with the Hamilonian.

We use a Diffusion Monte Carlo algorithm very similar to that described in Ref \cite{Foulkes_review}.
The Reptation Monte Carlo algorithm is from Ref \cite{pierleoni_rmc}, except that we
use the approximation to the Green's function as described in Ref \cite{Foulkes_review}.



Diffusion Monte Carlo and Reptation Monte Carlo both suffer from the sign problem
for fermions, which forces us to make the fixed-node approximation, where
the nodal surface of the exact wave function are assumed to be the same as the 
trial wave function.  This approximation typically results in recovering 
90-95\% of the correlation energy.  This and the pseudopotential localization approximation\cite{lubos_psp} are 
the only uncontrolled approximations in our calculations.  All calculations will be done using these two approximations.

We can control the fixed node approximation somewhat by varying the orbitals in the trial 
wave function and minimizing the DMC energy.  For transition metal oxides, this turns out to be
important, since Hartree-Fock orbitals are rather inaccurate and biased towards
more ionic picture of the state.  We have previously optimized the
mixing percentage in B3LYP in Ref \cite{CPL_lucas}, and it turns out that B3LYP orbitals are almost
optimal, so in these calculations we simply use the B3LYP orbitals.  We will also investigate using multiple determinants to improve the fixed node error for a case study of TiO.

\subsection{Bayesian Fitting of Bond Lengths}

The main disadvantage of QMC methods is that every quantity has a statistical uncertainty which decreases only as the square root of the computer time. For quantities like bond lengths, researchers have historically calculated the energy at several bond lengths, then fitted a function to the points.  Uncertainties have been calculated in many ways, but to our knowledge, none of them is exact and makes use of all the information available including the statistical uncertainty.  Here we offer a more systematic way of finding the minimum bond length along with its 
statistical error bar.

According to Bayes' theorem, given a model $M$ and a set of 
data $D$, the probability of the model given the set of data is:
 $P(M|D)=P(D|M)P(M)/P(D)$.  $P(D)$ is an unimportant normalization constant,
$P(M)$ is called the prior distribution, which we are free to set to reflect the 
{\it a priori} probability distribution on the set of models.  Without any good reason to believe otherwise, we 
generally set $P(M)=1$, the maximum entropy/least knowledge condition.  In the case of normally distributed data on a 
set of points $\lbrace x_1,x_2,...,x_N \rbrace $, 
\begin{equation}
P(D|M) \propto exp[-\sum_i (M(x_i)-D(x_i))^2/2\sigma^2(x_i)], 
\end{equation}
where $\sigma(x)$ is the statistical uncertainty of $D(x)$.

In the case of our bond lengths, we limited our space of models to $M(x)=c_1+c_2x+c_3x^2$, for $x$ close to 
the minimum bond length.  This is equivalent to setting the prior distribution equal to one for all quadratic
functions and to zero for non-quadratic functions.  We then calculated several data points $D(x)$ with statistical 
uncertainties $\sigma(x)$.  The probability distribution function of the bond length $b$ was then calculated by
evaluating the integral 
\begin{equation}
p(b)=\frac{\int \delta(-c_1/2c_2-b) P(M|D)P(M) dc_1dc_2dc_3}{\int P(M|D)P(M) dc_1dc_2dc_3}.
\end{equation}  This integral is only three-dimensional, and as such could be calculated by a grid method, but we found it convenient to calculate it by Monte Carlo, by sampling $P(M|D)P(M)$ and binning the bond length.  In all cases studied, $p(b)$ was very accurately a Gaussian distribution function, 
and so we report the bond lengths as an expected value plus a stochastic uncertainty, which fully
characterizes the distribution.

\subsection{Computational Parameters}

For the oxygen atom, we used the pseudopotential from Lester\cite{Lester_psp}, and for the transition metals, we used Ne-core soft potentials from Lee\cite{Lee_psp}.  To prepare the orbitals for the QMC calculation, we used 
GAMESS\cite{gamess} with a triple-zeta basis set.  Both RMC and DMC calculations were performed with $\tau=0.01$ Hartree$^{-1}$, which was converged within error bars, and for our RMC calculations, we chose $N=301$, which corresponds to a 3 Hartree$^{-1}$ long projection length. We evaluate the dipole moment within RMC as the expectation value of 
${\bf \mu}=e\langle \sum_ie{\bf r}_i \rangle +{\bf \mu}_{nuclei} $, using the Hellmann-Feynman theorem.

\section{Results and Discussion}

\subsection{Energetics}

\begin{figure}
\includegraphics[width=8.0cm]{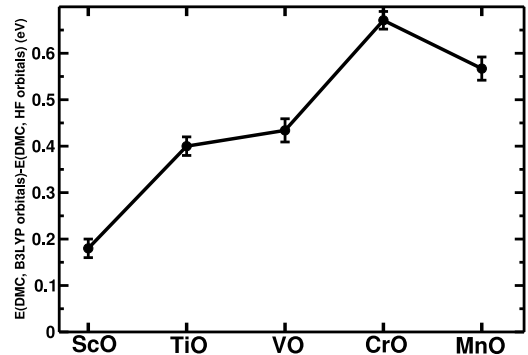}
\caption{The energy gain in DMC from using B3LYP orbitals as a function of the metal monoxide. The line is a guide to the eye. }
\label{fig:energy_gain_b3lyp}
\end{figure}

We begin with the importance of the one-particle orbitals used in the trial function.  These are not optimized
within VMC and the Jastrow factor does not change the nodes, so we are forced to use the nodes of the Slater determinant of orbitals from 
DFT or Hartree-Fock.  For systems without strong
electron correlation(for example, first and second row elements), it has been standard practice to use DFT and Hartree-Fock orbitals interchangeably, and the fixed-node
energy is fairly insensitive.  In TMO's, the correlation is much stronger and changes the structure of the one-particle orbitals.  
That is, the orbitals that define the lowest energy nodal structure are significantly different from the Hartree-Fock orbitals.
  Direct optimization of the orbitals within QMC is desirable, but very difficult for larger systems, so we took the approach of finding an optimal mean-field that produces orbitals that minimize the fixed-node energy.  In these systems, the hybrid functional B3LYP appears to be near-optimal.  In Fig \ref{fig:energy_gain_b3lyp}, we report on the energy gain in DMC by using B3LYP orbitals.  In our five molecules, there are roughly three levels of energy gain, corresponding to the type of bonding.  ScO 
has only one d electron in a $\sigma$ state, TiO and VO are respectively $\sigma^1\delta^1$ and $\sigma^1\delta^2$, and CrO and MnO are $\sigma^1\delta^2\pi^1$ and $\sigma^1\delta^2\pi^2$.  Each new type of symmetry adds approximately 0.2 eV to the energy gain in using B3LYP orbitals, with a slight decrease for the half-filled shell of MnO.  This energy gain is a measure of how poor the independent-electron approximation is for preparing the one-particle orbitals.  Since there is almost no gain in the atomic systems by using B3LYP orbitals, the correct orbitals appear to be critical for high accuracy in TMO materials, even more so as more d-symmetry electrons are present.

 We compare our QMC results to Density Functional Theory in the LDA, Coupled Cluster with singles, doubles, and perturbative triples(CCSD(T)), and a new hybrid meta-GGA, TPSSh, which should be the most accurate semi-empirical DFT available\cite{staroverov:12129}.  Using accurate one-particle orbitals, DMC binding energies(Table \ref{table:binding}) all fall within experimental uncertainties except for CrO
and MnO, which both have $\pi$-type electronic configurations.  The RMS deviations of DMC are around 50\% smaller than TPSSh and CCSD(T) at 0.21 eV.  This is still above the systematic error of 0.05 eV that would be required for `chemical accuracy'; however, the uncertainties of the experiments are also above this threshold.  

\begin{table}
\begin{ruledtabular}
\begin{tabular}{lcccccc}
Method & ScO & TiO & VO & CrO & MnO  & RMS \\
LDA\cite{furche:044103} & 9.09 & 9.13 & 8.48 & 6.26 & 6.51 & 2.19 \\
CCSD(T)\cite{baushlicher:189} & 6.71 & 6.64 & 6.13 & 4.20 & 3.43 & 0.31 \\ 
TPSSh\cite{furche:044103} & 7.11 & 7.18 & 6.44 & 4.45 & 4.62 & 0.38 \\
DMC &  7.06(3) & 6.81(3) & 6.54(3) & 3.98(2) & 3.66(3) & 0.21\\
Exp\cite{merer_review} & 7.01(12) & 6.92(10) & 6.44(20) & 4.41(30) & 3.83(8) & 0\\
\end{tabular}
\end{ruledtabular}
\caption{Binding energies of the first five transition metal monoxides
by different theoretical methods, along with RMS deviations from the experiment(all in eV). Statistical uncertainties in units of $10^{-2}$ eV are shown in parentheses for Monte Carlo and 
experimental results. Zero point energy corrections are estimated to be much less than the size of the uncertainty in experiment.  }
\label{table:binding}
\end{table}

Table \ref{table:bond_lengths} shows the calculated versus experimental bond lengths for the selected methods.  Here again, we see that DMC using a Slater determinant of B3LYP orbitals is quite accurate, with RMS deviation less than 0.01 \AA, close to the size of our statistical error.  It is again around 50\% more accurate than the high-accuracy methods CCSD(T) and TPSSh, and four times more accurate than the LDA on these systems, on average.

\begin{table}
\begin{ruledtabular}
\begin{tabular}{lcccccc}
Method & ScO & TiO & VO & CrO & MnO & RMS \\
LDA\cite{furche:044103} & 1.644 & 1.597 & 1.564 & 1.584 & 1.602 & 0.033 \\
CCSD(T)\cite{baushlicher:189} & 1.680 & 1.628 & 1.602 & 1.634 & 1.66 & 0.011 \\ 
TPSSh\cite{furche:044103} & 1.659 & 1.613 & 1.582 & 1.612 & 1.628 & 0.012 \\
DMC &  1.679 & 1.612 & 1.587 & 1.617 & 1.652 & 0.008 \\ 
Exp\cite{merer_review} & 1.668 & 1.623 & 1.591 & 1.621 & 1.648 & 0\\
\end{tabular}
\end{ruledtabular}
\caption{Bond lengths in \AA.  The statistical uncertainties for ScO,TiO,VO,CrO, and MnO
are respectively 0.002,0.003,0.003,0.004, and 0.004. }
\label{table:bond_lengths}
\end{table}

\subsection{Dipole Moment}

The dipole moment of these molecules has been noted as a difficult quantity to accurately calculate\cite{furche:044103,baushlicher:189}. In approaches relying on basis functions, there appears to be a large sensitivity to quality of the basis set.  It also appears to be very sensitive to an accurate treatment of the correlation.  The RMC method depends only very weakly on the basis set used 
to prepare the orbitals, and reaches the lowest energy of any variational method on these systems, 
so one may hope that RMC agrees with experiment better than other ab initio methods.
We find this not to be the case.  As shown in Table \ref{table:dipole_moments}, we find serious disagreement with experiment in three of the four molecules with experiments available.  Only ScO, the molecule with the weakest d-character, agrees well.  The rest are universally predicted to be much higher in QMC.

\begin{table}
\begin{ruledtabular}
\begin{tabular}{lcccccc}
Method & ScO & TiO & VO & CrO & MnO  \\
LDA\cite{furche:044103} & 3.57 & 3.23 & 3.10 & 3.41 & --  \\
CCSD(T)\cite{baushlicher:189} & 3.91 & 3.52 & 3.60 & 3.89 & 4.99 \\ 
TPSSh\cite{furche:044103} & 3.48 & 3.43 & 3.58 & 3.97 & --  \\
RMC &  4.61(5) & 4.11(5) & 4.64(5) & 4.76(4) & 5.3(1)  \\
Exp\cite{steimle_review} & 4.55 & 3.34(1)\cite{steimle_tio_03} & 3.355 & 3.88 & -- \\
\end{tabular}
\end{ruledtabular}
\caption{Dipole moments in Debye.  The fixed-node RMC results have been obtained with a 
single deteriminant of B3LYP orbitals.  See text for an analysis of the errors involved
for the case of TiO.}
\label{table:dipole_moments}
\end{table}

The significant discrepancies from experiment are surprising given the excellent agreement that we obtained with energies.  It also seems strange that the LDA is generally quite close to the experiment, since we know that for energetics it performs relatively poorly.  We analyze the errors present in the calculations as follows for the case of TiO, the simplest of the molecules with a large difference from experiment.
\begin{itemize}
\item {\bf Pseudopotential error}. We checked the pseudopotential in mean-field calculations, and 
it caused an overestimation of the dipole moment in TiO by 0.1 Debye.  This is systematic for all five materials studied,
with each having an overestimation of between 0.1 and 0.15.
\item {\bf Hellmann-Feynman theorem}.  The definition of dipole moment is ${\bf \mu}=\frac{d\langle H \rangle}{d{\bf E}}$, where
${\bf E}$ is the electric field.  We have used the Hellmann-Feynman theorem to instead evaluate it as 
${\bf \mu}=e\langle \sum_ie{\bf r}_i \rangle +{\bf \mu}_{nuclei} $. As shown in Ref \cite{huang:4419}, the Hellmann-Feynman theorem for
calculating the dipole moment does not exactly apply in fixed-node Quantum Monte Carlo, although the errors are 
generally considered small.  To check this, we calculated the dipole moment using the finite field approach and correlated sampling\cite{filippi_force} using nodes from B3LYP also at that field, and obtained the same result as the Hellmann-Feynman estimator within error bars.  
\item {\bf Fixed-node error and localization}.  The only remaining approximation in our simulation is the combined fixed-node error and localization error.  We investigate this by
attempting to systematically improve the trial wave function.
\end{itemize}
To go beyond the Slater-Jastrow form, we began to add more determinants into the trial function for the test case of TiO.  We
performed a configuration interaction calculation including single and double excitations starting from the B3LYP orbitals, kept the determinants with the highest weights, and reoptimized the weights in the presence of the Jastrow factor.  If we did not reoptimize the weights, we found that the fixed-node energy actually {\it increases}, suggesting that the correlation present in the Jastrow factor is critical to an accurate description of these materials.  The Jastrow factor also reduces the number of determinants necessary to describe the electron correlation by several orders of magnitude, since the Jastrow factor contains most of the so-called dynamic correlation.  Finally, we used an RMC simulation to find the dipole moment using the Hellmann-Feynman theorem.  
\begin{figure}[h]
\includegraphics[width=\columnwidth]{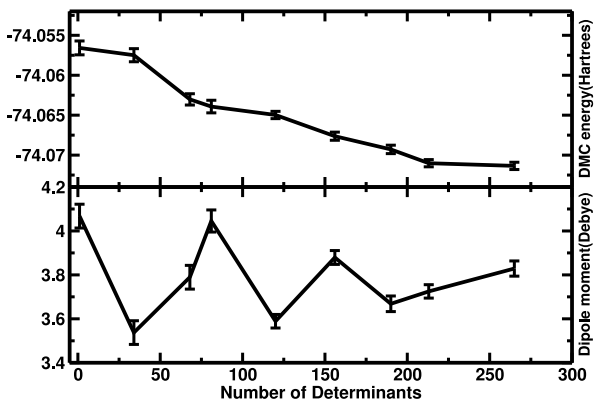}
\caption{The number of determinants versus the energy and dipole moment for TiO.The dipole moments are shifted downwards by 0.1 Debye to correct for the pseudopotential error. }
\label{fig:tio_multidet}
\end{figure}
Fig \ref{fig:tio_multidet} shows significant improvements in the energy on the order of .015 Hartrees, and the dipole moment
appears to oscillate around a value of approximately 3.9(1) Debye.  
Therefore, we estimate the error in dipole moment to be approximately 0.2 Debye from the fixed-node approximation and 
0.1 Debye from the pseudopotential approximation.
With these corrections, the minimum value of the dipole moment for TiO is 3.6 Debye with over 95\%
confidence.  This is consistent with the CCSD(T) number, but still visibly
 larger than the value reported by experiment.

\section{Conclusions}

We have found that for energetics, DMC using a single determinant trial function is remarkably accurate, 
perhaps suggesting that for 
these materials, it is sufficient.  The bond lengths and binding energies are, on average, 50\% better than the best meta-GGA and CCSD(T).  
  The Bayesian method for finding the minimum bond lengths mitigates the inconvenience of statistical uncertainty, while improving the performance by using all possible information.

The dipole moment appears to be more challenging, and requires a complicated treatment of the 
wave function nodes to obtain a stable value with respect to changes in the nodes.  We have obtained
a converged value for TiO, however, and it is still somehwat higher than suggested
by experiment.  This is in line with 
the large values for the dipole moment obtained by CCSD(T) and B3LYP;  
agreement with CC method is particularly reassuring.
 In addition, there are to our knowledge only two experimental measurements
of the dipole moment\cite{steimle_tio_03,steimle_89} from the same group, which report significantly
different moments(2.96 versus 3.34 Debye).  This may indicate a sizeable uncertainty in the experiment as well.  
The apparent agreement of LDA is almost certainly fortuitous, because LDA underestimates the
bond length, which in turn causes the dipole moment to be too small.  In fact, one would 
generally expect LDA to underestimate the dipole moment even at the correct bond length, 
since it tends to make the molecule not ionic enough.  This may also indicate an inaccuracy 
in the experiment.  One would expect that the corrections for the fixed-node approximation for
VO, CrO, and MnO are similar or slightly greater than TiO, and therefore are around 0.2-0.4 Debye.
It is well-known that dipole moment is particularly sensitive to contributions from single excitations
which, however, have only very minor contributions to energy by coupling through doubles. It is therefore
plausible that the optimization procedures which we have employed were not able to recognize
the weak signal from singles and therefore the wavefunctions are still not perfect in this
respect.
The dipole moment remains an extraordinarily sensitive quantity that is a stringent test of 
theory and experiment.  It may be interesting to see if there are other wave functions that can describe the
nodal surface to sufficient accuracy while being more compact than a large determinantal expansion.


We would like to acknowledge the NSF Graduate Research Fellowship program, NSF  DMR-0121361, and EAR-0530110 grants
 for support of this work and the North Carolina 
State Physical and Mathematical Sciences cluster and NCSA for providing compute time.
\bibliography{tmo}
\nocite{*}

\end{document}